# Performance of a crystalline silicon photovoltaic power plant during sandstorms

B. Ravindra

*Abstract*— Solar photovoltaic power generation has achieved grid parity in a number of countries during the last decade. This revolution has begun in countries where the solar radiation intensity is not very high. Solar resource maps created by a number of agencies worldwide indicate that areas of high intensity of solar radiation suitable solar for power generation are in deserts and semi-arid zones. These regions are dusty and are subjected to both anthropogenic pollution and natural sand storms. The impact of these events on the solar photovoltaic (PV) plant yield is significant. An increase in diffuse component of solar radiation is accompanied by a naturally occurring sand storm. Often the solar panels are tilted at an optimal angle and the global tilted radiation value falling on the panels changes significantly during and after the sand storms. It is noted that the decrease in the power output of a solar photovoltaic plant during such events is high enough to warrant further consideration while forecasting the plant output. Inverter performance during such sand storms is found to be satisfactory. Integrating dust storm monitoring with PV power plant yield prediction algorithms is necessary for an improved operation of PV plants.

*Index Terms*—Solar radiation, solar photovoltaic power plants, sand storms

## I. Introduction

Renewable energy sources such as wind, solar, geothermal and biomass have received considerable attention during the last two decades. The rise in oil prices (in 2007-2008) and a significant decrease in the cost of PV module prices have helped wide spread deployment of solar energy. Both solar photovoltaic (PV) and solar thermal power plants have been set up in various parts of the world to augment the conventional power generation. Rooftop solar PV power plants have led the solar growth in developed world. But utility scale large solar PV power plants have contributed to the growth of solar in developing countries. Global Horizontal Irradiance (GHI) component of solar radiation is an important factor to be considered in setting up solar PV power plants. But in regions where sand storms are common the diffuse part of the solar radiation also needs to be considered. A detailed investigation of diffuse and global components of solar radiation and their variability during sand storms helps in improved solar resource assessment. These considerations may also lead to new algorithms for forecasting power output from solar power plants.

Solar resource assessment traditionally considered the use of typical meteorological year data. Nowadays, good quality satellite data is also available from multiple sources. However, good quality ground data is often necessary for final decision making. The quality of satellite data for annual average values of GHI is approaching that of the ground station data for a standard pyranometer. There is also a need to understand the individual effect of aerosols and clouds on solar radiation for better forecasting of solar power output. It is sometimes necessary to distinguish the aerosol contribution from anthropogenic pollution (vehicular, industrial emissions and biomass burning) from naturally occurring sand storms. This aspect has received considerable attention from environmental researchers studying air quality but has not yet become a tool for solar power plant resource assessment and forecasting. It may be noted that the implications of solar dimming and brightening phenomenon (studied by climate scientists) for solar photovoltaic power plants have also been examined recently [1]. The projected change in the power output of PV plants has been shown to be not very significant. They concluded, "Despite small decreases in production expected in some parts of Europe, climate change is unlikely to threaten the European PV sector". These studies deal with long term averages and trends but a solar plant operator often needs to deal with short term forecasting. The impact of climactic events in such cases is very significant and changes in solar radiation data during these events impact solar resource assessment and forecasting. Integrating cloud movement vectors using sky imager data in real time with solar plant forecasting has also received attention [2-4]. Similarly, there is a need to integrate dust storm warning methodology with solar power plant yield forecasting.

The implications of soiling of panels on PV power plant yield have been examined by a number of researchers. These range from recommendations of de-rating factors as proposed by Sukhatme and Nayak [5] to detailed investigations based on the nature of dust deposited [6]. Often these studies involve recommendations of PV panel cleaning schedules. A detailed study of soiling loss on photovoltaic modules with artificially deposited dust of different gravimetric densities and compositions collected from different locations in India has been carried out by John et al [6]. Influence of dust deposition on photovoltaic panel performance has been carried out by

This paper is submitted on 30th November 2017. This work was supported in part by the MHRD & MNRE and IMD, Pune, Government of India.
B. Ravindra is with the Department of Mechanical Engineering, Indian Institute of Technology, Jodhpur, Nagaur Road, Karwar, Rajasthan, 342 037 INDIA (e-mail: ravib@ iitj.ac.in).

Abhishek Rao [7]. Performance evaluation for PV systems to synergistic influences of dust, wind and panel temperatures from a spectral point of view has been studied by Khanum et al [8]. An industry perspective on soiling and its effect on PV panels can be found in Atanometric document [9]. Implication of dust deposition for degradation of PV panels and field level issues in the Indian context has been carried out by Rajiv Dubey et al [10]. Availability of automated cleaning systems for PV plants alleviates the soiling problem to a certain extent. The current study differs from these soiling studies in that the dynamic performance of a PV plant is studied during a sand storm event. The dynamic changes in the solar radiation due to increased aerosol content can impact the power output. This aspect has been considered in the current work. This issue has implications for short term forecasting of PV power plant yield and also for the inverter performance during such events. Often such dust storms may be followed a drop in temperature and a good shower of rain (which may be enough to clean the panels). But if the rainfall does not accompany the dust storm then its impact is more significant.

The impact of dust storms on people living in the arid and semi-arid zones has received wide attention. Most of these investigations have studied health and agricultural and transportation aspects. It is known that the dust storms from Sahara desert can impact cities in North America and sand storms in Middle East can continue towards south Asia. The impact of these events on the solar power generation needs detailed investigation as well. The Desertec project in Sahara is an example in this regard. Several solar projects in the MENA area also are influenced by sand storms. This study covers a case of a sand storm that has its origins in Middle East and has reached Northern and Western parts of India during $18^{th}$-$23^{rd}$ March 2012 [11]. The solar radiation data obtained from a ground station at Jodhpur in India (maintained by Indian Meteorological Department (IMD)) and its impact on solar power generation due to this storm is presented. India has currently 12 GW solar power plants and most of these are located in the western Rajasthan and Gujarat which are impacted by such sand storms. Detailed understanding of sand storms can assist in solar resource assessment for setting up future power plants and their grid integration. These results can also be extended to include other regions where similar events occur [11-14].

This article considers the case of a roof top PV plant performance in Jodhpur, during a sand storm. The state of Rajasthan is one of the prime hubs of solar power station in India. This roof top plant is established 2011 to study field performance of solar PV panels under desert conditions. Specific objectives included are (i) to study degradation aspects of solar panels, (ii) benchmarking yield assessment software with site specific data, (iii) use it as a training platform to develop skilled human resource in solar energy, (iv) to understand the grid integration issues in a context where grid reliability is a matter of concern. In this article, the effect of a sand storm on this rooftop PV plant is presented.

## II. DESCRIPTION OF THE DUST STORM EVENT

Sand storms occur in Thar desert region in India during summer and pre-monsoon season. The frequency of such events varies from year to year and may reach up to ten in one calendar year. The city of Jodhpur (Latitude $26.2^0$N and Longitude $73^0$E and 290m above MSL) witnesses every year several such storms during the March-June months. The Indian Meteorological Department (IMD) has a weather station at Jodhpur that has capability to monitor solar radiation and other weather related parameters to monitor sand storms. Detection of these events based on observations from microwave satellites, visible and infrared instruments as well as TOMS Aerosol index is presented by IMD experts [11]. These results are also correlated with results obtained from sky radiometer at the Jodhpur ground station. They [11] wrote "On March 20, 2012, a giant dust plume stretched across the Arabian Sea, from the coast of Oman in the west to the coast of India in the east. This extensive plume followed days of dust-storm activity over the Arabian Peninsula and Southwest Asia, including north and northwest part of India. Gulf News reported that several meteorologists had characterized the late March dust activity in this region as a "super sandstorm" with effects reaching as far as Southeast Asia. The dust storm resulted from two different storms converging. The first front carried dust from Iraq and Kuwait, and the second front stirred dust in southeastern Iran." This dust storm persisted till $23^{rd}$ March 2012 as per IMD report [11].

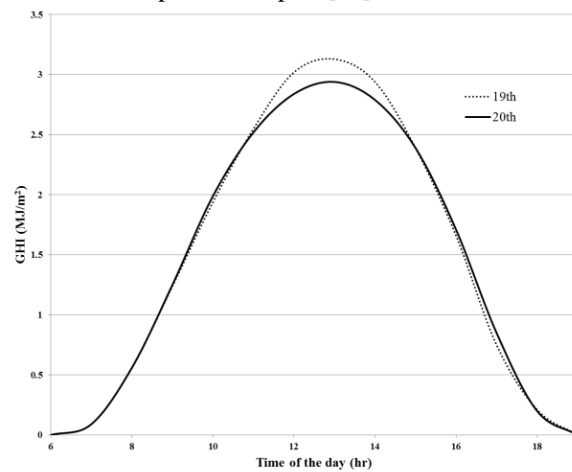

Fig 1. GHI (MJ/m$^2$) on $19^{th}$ and $20^{th}$ March 2012

It is interesting to investigate the solar radiation components and their variability during this event. Towards this end, the hourly averaged values of two components of solar radiation are considered in this work during these days: One day before the dust storm, during the dust storm (on $19^{th}$ March 2012) and one day after the dust storm ($20^{th}$ March 2012). The hourly average values of Global Horizontal Irradiance (GHI) for these days ($19^{th}$ and $20^{th}$ March 2012) are shown in Fig 1. It may be noted that 1 MJ/m$^2$= 277.8 Watt-h/m$^2$. As expected, the GHI value has decreased during the dust storm. The daily average GHI value decreased from 200 W/m$^2$ on $19^{th}$ March 2012 to 190 W/m$^2$ on $20^{th}$ March 2012 (2% decrease from previous day). The peak value decreased from 400 W/m$^2$ to 380 W/m$^2$ (6% decrease from previous day). The hourly averaged diffuse component is shown in Fig 2.

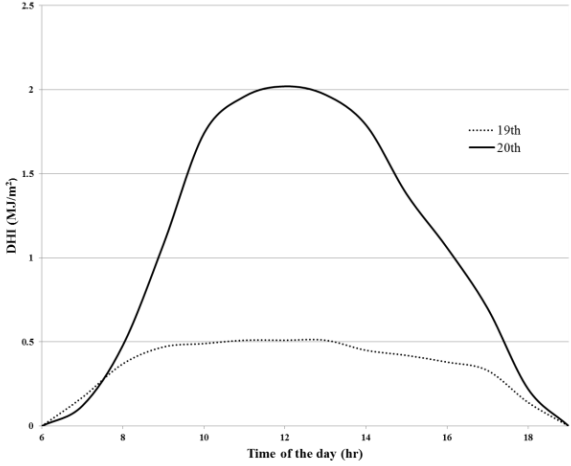

Fig 2. DHI (MJ/m$^2$) on 19$^{th}$ and 20$^{th}$ March 2012

It can be observed that due to increased aerosol content during the dust storm, the DHI value increased substantially on 20$^{th}$ March 2012 (250% increase in peak value from previous day). It continued to be high even the day after the dust storm (unlike the GHI value). Most of the dust storms are also associated with cirrus clouds. The effect of such clouds and aerosols on PV plant performance needs further study. For solar resource assessment and performance evaluation of PV plants, only the GHI component of solar radiation is considered in many instances. The diffuse component (DHI) is often considered not very relevant for solar PV power generation. Thus a 2% change in the average value of GHI (and 6% change in the peak value of GHI) ought not to have large impact on the yield of the PV power plant. This is considering the fact that the short circuit current of the PV cell is almost proportional to the radiation. However, it has been observed that there is significant drop in the PV plant power output during the dust storm. The reasons for this are explored in this article through a detailed study of the solar radiation on the tilted surface and by examining the measured values of solar PV cell radiation sensor (also positioned at the same angle as the PV panel).

It is often the case that the PV panels are oriented at a suitable angle to maximize the incident solar energy. In such situations it is necessary to compute the Global Tilted Irradiance (GTI or $I_T$). Hourly averages of tilted global tilted irradiance, global horizontal irradiance and diffuse horizontal irradiance are denoted by $I_T$, $I_g$, $I_d$ respectively and are related by the following equation [5]:

$$\frac{I_T}{I_g} = r_b \left[1 - \frac{I_d}{I_g}\right] + r_d \left[\frac{I_d}{I_g}\right] + r_r \quad (1)$$

Where $r_b$, $r_d$, $r_r$ are the tilt factors for beam radiation, diffuse radiation and reflected radiation respectively. These are given by the following expressions:

$$r_b = \left[\frac{\cos(\theta)}{\cos(\theta_z)}\right] = \left[\frac{\sin(\delta)\sin(\emptyset-\beta)+\cos(\delta)\cos(\omega)\cos(\emptyset-\beta)}{\sin(\emptyset)\sin(\delta)+\cos(\emptyset)\cos(\delta)\cos(\omega)}\right] \quad (2)$$

$$r_d = (1 + \cos(\beta))/2 \quad (3)$$

$$r_r = \rho(1 - \cos(\beta))/2 \quad (4)$$

Here the latitude (φ), hour angle (ω), angle of declination (δ) and solar zenith angle (θ$_z$) depend on the site location, time and day of the year [5]. The angle of tilt (β) of the PV panel is taken to be 25$^0$ and the reflectivity of the ground (ρ) is assumed to be 0.2. It may be noted that the tilt factor for the diffuse radiation is based on the assumption that the sky is an isotropic source. There exist other models which account for the anisotropy of distribution of diffuse radiation. These are not considered in this article. The assumption of isotropy may suffice for the purpose of predicting the yield of the PV plant. The computed value of GTI (or $I_T$) is shown in the Fig. 3. It is clear that due to the large decrease in DHI, the GTI value has reduced substantially when compared with GHI (shown Fig. 1). Thus, it is necessary to account for the tilted radiation value as the diffuse component variation gets reflected in the overall radiation available for power generation. It may be noted that in Fig. 3, the drop in peak value of GTI from previous day is 20%. Hence, the power drop due to the dust storm event is expected to be of the same order. In the next section, a description of the PV plant is given and the actual power reduction measured is compared to the predicted value due to the drop in GTI. Thus, forecasting the decrease in GTI value can help in predicting the power drop due to such extreme events.

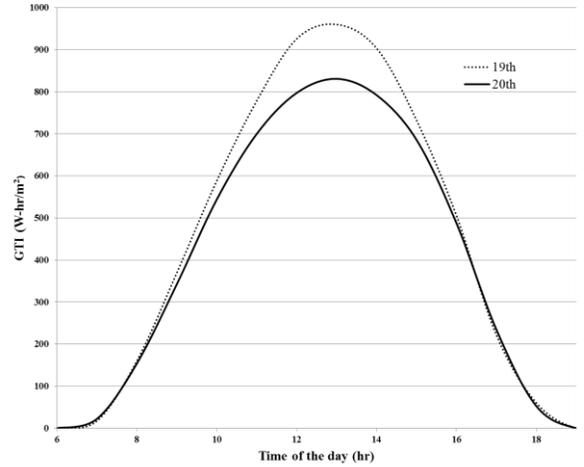

Fig 3. GTI on 19$^{th}$ and 20$^{th}$ March 2012

### III. DESCRIPTION OF PHOTOVOLTAIC PLANT

The photovoltaic plant considered in this article has a total installed capacity of 58 kW and consists of six power conditioning units (PCU). Each PCU of 10 kW is a Sunny Mini central 1000TL model from SMA, Germany. There are 18 strings and each string consists of 15 PV Crystalline Silicon (C-Si) panels (from Moser Baer Photovoltaic Limited, India) of rated capacity 215 Wp in series. The module electrical specifications are as follows: Maximum Power ($P_{max}$): 215 W; Voltage at $P_{max}$ ($V_{mp}$): 29.21 V; Current at $P_{max}$ ($I_{mp}$): 7.36 A; Open Circuit Voltage ($V_{oc}$): 36.21V; Short Circuit Current ($I_{sc}$): 7.93A; The temperature coefficients of the PV panel are given by (i)Temperature Coefficient of $P_{max}$(%/K): -0.43; (ii)Temperature Coefficient of $V_{oc}$(%/K): -0.344; (iii) Temperature Coefficient of $I_{sc}$(%/K): 0.11. Total number of cells per by-pass Diode is 20 in number. The arrangement and number of cells is given as 156mmx156mm

Multicrystalline Silicon Solar PV Cells, with a 6x10 configuration. Dimensions of each module are 1661mmx 991mmx40mm and its weight is 19.5 Kg. Total number of modules in this grid tied roof top PV plant is 270.

The plant is also equipped with a cell radiation sensor (which is placed at the same angle as PV modules), ambient as well module back sheet temperature monitoring sensors. The data acquisition system is also supplied by SMA and is referred to as Sunny web box. The data from the plant can be accesses anywhere from the sunny web portal. The PV cell radiation sensor data is shown in Fig. 4. It can be seen that the drop in the peak value is around 21%. This is similar to that computed for GTI in Fig. 3. Thus, in the absence of measured values of DHI and GHI (for which a shaded ring pyranometer and a pyranometer are respectively required), a cell radiation sensor may suffice. This is a cheaper alternative though it may not capture the entire spectral content of solar radiation incident on the module. How the spectral changes occur during the dust storms due to increase in the diffuse component has not been thoroughly investigated in the literature.

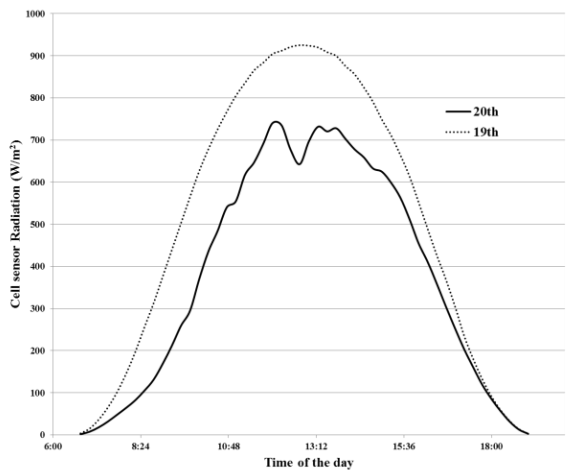

Fig 4. Cell sensor data on 19$^{th}$ and 20$^{th}$ March 2012

The output from one of the strings during 19$^{th}$-21$^{st}$ March 2012 is shown in Fig.5. The PCU operates in Maximum Power Point Tracking (MPPT) mode with a high efficiency of 97-98%.

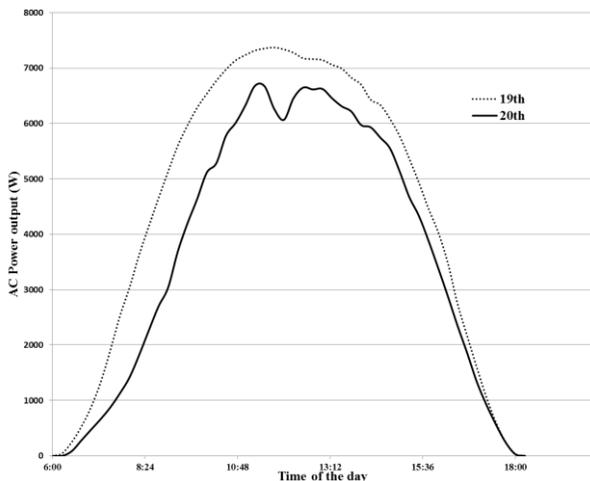

Fig 5. AC Power output (W) on 19$^{th}$ and 20$^{th}$ March 2012

The short circuit current from the PV cell and the ac current are shown in Fig. 6. This clearly shows that the plant did not trip during the event and that the PCU functioned as expected.

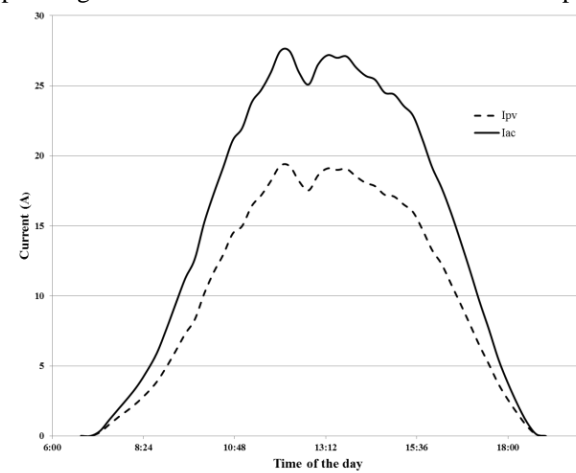

Fig 6. DC and AC current output on 20$^{th}$ March 2012

It is instructive to examine the maximum power output from a PV cell and correlate the effect of dust storm in that context. These results can explain qualitatively the observed features in the output of the plant. Using a simple equivalent circuit model, the current in a PV cell can be described as the maximum power output from a PV cell ($P_{max}$) at a given cell temperature $T$ can be derived [5] as given by the following equation:

$$P_{max} = \frac{\frac{eV_{mp}^2}{kT}}{\left(1+\frac{eV_{mp}}{kT}\right)}(I_{sc}+I_0) \quad (5)$$

Where, $e$ is the charge of an electron, $k$ is the Boltzmann constant, $T$ is the absolute temperature of the cell ($^0$K), $I_{sc}$ is the short circuit current and $V_{mp}$ is the voltage at which the power is maximum. $I_0$ is the reverse saturation current can be calculated [5] using the following equation:

$$I_o = \frac{I_{sc}}{\left(\exp(\frac{eV_{oc}}{kT})-1\right)} \quad (6)$$

The changes in short circuit current can be considered as proportional to changes in solar radiation (GTI). Hence a decrease in solar radiation due to the dust storm results in a proportional decrease in short circuit current which in turn causes the power from the cell to drop. The current in the cell can be calculated [5] using the following equation:

$$I = I_{sc} - I_o[\exp(\frac{eV}{kT})-1] \quad (7)$$

Thus it can be seen that knowing the cell temperature is critical for determining the performance of the PV cell. It is known that a dust storm event is often accompanied by a drop in ambient temperature and increase in wind speed. Both these factors contribute to a decrease in cell temperature. Given temperature coefficients of the PV cell, it is possible to quantitatively examine the effect of this phenomenon. An empirical expression for the PV panel back surface module temperature ($T_m$ in $^0$C) is given in Sandia lab report as follows:

$$T_m = E\{e^{a+bW}\} + T_a \quad (8)$$

Where $T_a$ is ambient air temperature ($^0$C), $E$ is Solar irradiance incident on module surface (W/m$^2$), W is the wind speed measured at standard 10 m height (m/s), a & b are empirically measured coefficients for various panel front and back covers. It may be noted that there are other empirical models to predict the module back surface temperature [15]. Once the module back surface temperature is known, then the cell temperature can be estimated using empirical correlations. It is known that cell temperature is often one or two degrees higher than the back surface temperature. The measured ambient and module back surface temperatures are shown in Fig. 7 for two days before and during the dust storm. As expected, due to the increased wind speed during the storm, the drop in module temp is higher than the drop in the ambient temperature. However, there is a net drop in the power output as the drop in the solar radiation plays a major role than the drop in module temperature. It is also important to note that temperature variation may not be uniform from one module to another due to the variation in convective heat transfer coefficients. Hence it may not be possible to extrapolate data from one module for the entire plant. Thus equivalent circuit based model predictions are useful for obtaining base line data for performance prediction of solar PV plants but are not a substitute for the popular statistical methods based on time series analysis.

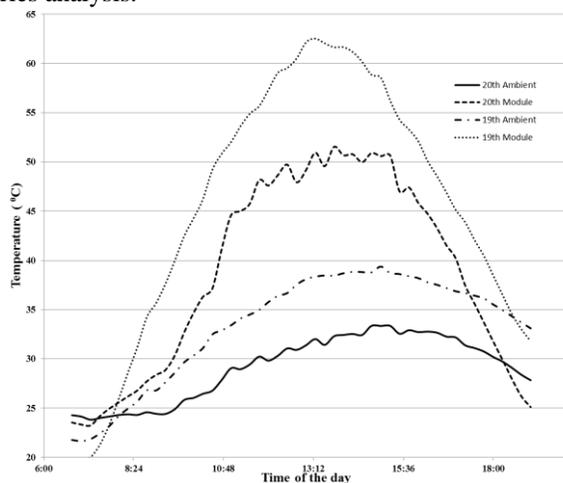

Fig 7. Temperature on 19$^{th}$ and 20$^{th}$ March 2012

Performance ratio is a standard metric to evaluate/assess the PV plants; PR= (Actual energy output from the plant)/ [(Incident solar radiation flux)x(Total area of the PV modules in the plant)x(PV module efficiency)]. It accounts for the losses in the process of producing electrical output. It is seen that that percentage drop in the PR for the plant from 19$^{th}$ March 2012 to 20$^{th}$ March 2012 is 9.4%. The capacity utilization factor (CUF) is another index that if often used in assessing PV plants in comparison with other technologies. It is defined as the ratio of actual PV plant yield in a given duration to the production of the PV plant at installed STC (Standard Test Conditions) capacity during the same duration. It is seen that the CUF has reduced from 25.25% on 19$^{th}$ March 2012 to 20.67% on 20$^{th}$ March 2012 due to the storm event. The drop in PR is higher than that of the CUF as PR takes into the effect of changes in radiation components due to the dust storm. This sudden drop in power generated has implications for operation and power evacuation schedules in case of grid connected plants.

IV. INVERTER PERFORMANCE

Inverter performance in desert areas is of great concern. A product brief from the inverter manufacturer SMA [16] reports that sand and dust tests were conducted on central inverters used in large scale in the environmental simulation laboratory at RUAG Land Systems AG (Thun, Switzerland), in accordance with IEC 68-2-68/EN 60068-2-68. According to this report, pulverized roof tiles whose composition was found to be similar to the sand found in the Arizona desert is used in this test. It is also mentioned that the dust particle concentrations (brick powder with a diameter of 7-20 microns and quartz sand with diameter between 0.1 and 0.6 millimeters) used in this testing exceed the limits established under the classification of ambient conditions according to IEC 60721-3-4, category 4S2, by several times. The wind speeds considered are of the order of 20 m/s. It is not clear whether the inverters used in rooftop installations are subjected to same type of testing. It may also be noted that the warranty offered in the case of large scale plants may be up to 20 years (sometimes matching that of PV modules). But for rooftop installation cited in this article, the warranty period for inverters is only five years. Plots of inverter efficiency (define as AC power output/ DC power input) during the sand storm event are shown in Fig.8.

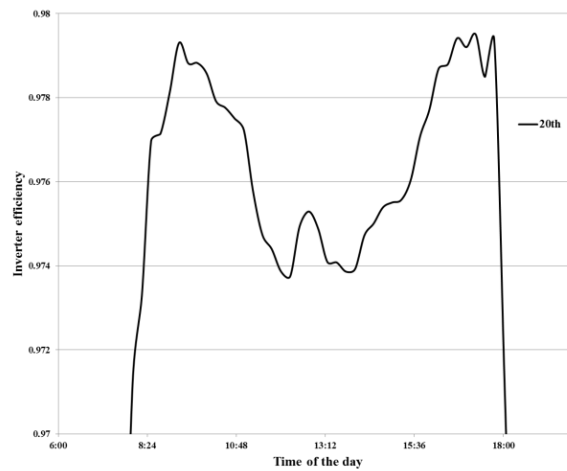

Fig 8. Inverter efficiency during the storm

It can be seen that no tripping of the plant has occurred during this event and the inverters seem to have performed well in tracking the changes in solar radiation due to the sand storm. Data at a much higher sampling rate than used here is required for further analysis. The frequency of dust storms in Jodhpur is about ten per year and the Thar desert is considered as one of the most populous deserts in the world. Thus both anthropogenic pollution and natural dust due to sand storms are to be considered in this context. Dust particles of small size are blown off the PV modules and inverters by strong wind currents but the larger particles tend to stay, contributing to loss of power. Sometimes dust storms are followed by a rain fall (shower). In such cases the power reduction on following days due to dust accumulation may not be significant. An examination of GHI and DHI components of the solar radiation during all the five days of the super sand storm (18$^{th}$ March to 23$^{rd}$ March 2012) indicate that though the changes in GHI are not that high, DHI decreases dramatically after the storm (20th March 2012) to limp back slowly to the clear sky values on 23rd March 2012. The decrease in yield of PV plants during all these days is rather significant (Daily yield being much below the expected yield for a clear sky day such as 19th March 2012). Thus one needs to track the dust/aerosol content in the atmosphere. Washing schedule should ensure that the PV panels are cleaned soon after the dust storms to minimize the power loss.

## V. Conclusions

In countries where Grid code is already in force, forecasting of grid tied PV power plant yield is necessary. Dust storms adversely impact the yield of solar PV power plants [17]. Incorporating the dust storm warning features in PV power plant forecasting software can help the utilities plan their operations effectively. In some regions around ten dust storms occur per year. They can significantly change the PV plant predictions and warrant further investigations. One can also learn from the Mars rover missions in this context.

Additional observations on amorphous silicon PV plants indicate that the drop in output due to the dust storm is much higher than that of the crystalline silicon PV plant. Thus, the increase in the diffuse component of solar radiation and the corresponding changes in the spectral content impact solar cells and modules differently. This aspect needs further study, though crystalline silicon seems to be dominant choice among solar power plant developers.

## Acknowledgment

The author thanks all the faculty members and students of the center for energy at IIT Jodhpur as well as Moser Baer photovoltaic limited personnel involved for their contribution in establishing the rooftop PV plant at IIT Jodhpur whose data is used in this article. He also thanks IMD, Pune, India for supplying the solar radiation data for Jodhpur.